\documentclass[prl,reprint,nofootinbib,superscriptaddress,preprintnumbers]{revtex4-1}
\usepackage{amsfonts}
\usepackage{amssymb}
\usepackage{amsmath}
\usepackage{graphicx,color}
\usepackage{dcolumn}
\usepackage{epsfig}
\usepackage{bm}
\usepackage{bbm}
\usepackage{ulem}

\usepackage{color}

\def\gtap{\ \raise.3ex\hbox{$>$\kern-.75em\lower1ex\hbox{$\sim$}}\ }
\def\ltap{\ \raise.3ex\hbox{$<$\kern-.75em\lower1ex\hbox{$\sim$}}\ }
\begin{document}
\preprint{LFTC-18-10/31}

\title{
Extraction of $\gamma n\to\pi N$ observables
from deuteron-target data
}
\author{Satoshi X. Nakamura}
\affiliation{
Laborat\'orio de F\'isica Te\'orica e Computacional - LFTC, 
Universidade Cruzeiro do Sul, S\~ao Paulo, SP 01506-000, Brazil
}

\begin{abstract}
An examination is conducted on
a commonly used procedure for
extracting (un)polarized $\gamma n\to \pi^-p$ and $\gamma n\to \pi^0n$
observables from $d(\gamma,\pi^-)pp$ and $d(\gamma,\pi^0)pn$ data, 
using a model that consists of 
the impulse term and the final-state
interaction (FSI) terms due to nucleon- and pion-exchange.
Recent experimental and theoretical analyses
used an extraction method that does not 
impose a cut on the final $\pi N$ invariant mass $W$. 
I demonstrate that the use of 
this method can result in 
the $\gamma n\to \pi N$ observables that are seriously distorted by 
the nucleon Fermi motion, 
and that one can efficiently avoid this problem by imposing a cut on
 $W$.
It is also shown that
the use of kinematical cuts of recent experimental analyses
can still leave in the selected samples substantial FSI effects
that must be corrected in extracting the $\gamma n \to\pi N$
cross sections.
In terms of the nucleon- and pion-exchange mechanisms,
I give the first qualitative explanation of
the FSI corrections, obtained in a recent MAMI experiment, 
for extracting $\gamma n\to \pi^0 n$ cross sections.
\end{abstract}


\maketitle

Extracting (un)polarized cross sections for pion photoproduction off the neutron,
$\gamma n \rightarrow \pi^- p$ and $\gamma n \rightarrow \pi^0 n$, 
from the deuteron-target data, 
$d(\gamma,\pi^-)pp$ and $d(\gamma,\pi^0)pn$, is an important task
at photon facilities such as 
Jefferson Laboratory (JLab)~\cite{jlab1,jlab2,clas1,clas2}
and MAMI~\cite{kossert,mami3,mami1,mami4,mami5,mami2,mami6},
forming a base for studying the baryon spectroscopy.
A commonly used procedure of 
extracting the $\gamma$-$n$ cross sections
is to apply a certain set of kinematical cuts to the
deuteron data and assume that the selected events are from
single-nucleon quasi-free processes.
For an accurate extraction, however, one may wonder
what corrections are needed (or not) to account for 
final state interaction (FSI) effects remaining in 
the selected events.
In the MAMI analysis~\cite{mami1,mami6}, the FSI corrections
for $\gamma n \rightarrow \pi^0 n$ cross sections
were assumed to be the same as those for $\gamma p \rightarrow \pi^0 p$,
which still needs a validation.
Also, the validity of the applied cuts is a question.
A theoretical analysis might answer these questions.

Tarasov et al. conducted 
a series of theoretical studies~\cite{jlab1,tara,tara-1} on
the FSI corrections needed to extract 
$\gamma$-$n$ cross sections.
Their model for $d(\gamma,\pi)NN$ is equipped with
two-body $\gamma N\to\pi N$, $\pi N\to\pi N$ and $NN\to NN$
elementary amplitudes generated with the SAID model~\cite{said},
and the off-shell momentum dependence of
the $NN\to NN$ amplitudes (the other amplitudes) is assumed to be 
a monopole form~\cite{lev06} (constant at the on-shell values).
They considered
the impulse, $N$-exchange, and $\pi$-exchange mechanisms
for $d(\gamma,\pi^-)pp$~\cite{jlab1,tara}, 
and estimated FSI corrections for extracting
$\gamma n \rightarrow \pi^- p$ cross sections.
The FSI corrections were then used in JLab
analyses~\cite{jlab1,clas2}.
However, their extraction formula
[Eq.~(\ref{eq:dsig_extract_mod2}) below]
neglects a possible effect from the nucleon Fermi motion, which needs a validation.
The authors also analyzed $d(\gamma,\pi^0)pn$~\cite{tara-1}
without including the $\pi$-exchange mechanism which was assumed 
to be negligible.
Their predicted FSI corrections for $\gamma N\to\pi^0 N$ cross sections turned
out to be even qualitatively different from what has been found in the
MAMI analysis~\cite{mami1,mami6}, which clearly calls for a further study.

In this Rapid Communication, 
I critically examine the extraction formula
[Eq.~(\ref{eq:dsig_extract_mod2}) below]
used in the JLab~\cite{jlab1,jlab2,clas1,clas2}
and theoretical analyses~\cite{tara,tara-1}.
I point out 
that the neutron-target observables extracted with this
formula can be seriously distorted by
the Fermi motion.
My calculation also finds significant
FSI corrections,
corresponding to kinematical cuts of the recent JLab analyses~\cite{jlab1,clas2,clas1},
needed to extract the neutron target observables.
For the first time,
I show that the FSI corrections,
obtained in the MAMI analysis of $d(\gamma,\pi^0)pn$~\cite{mami1,mami6},
for $\gamma N\to \pi^0 N$ cross sections
are reasonably well explained by 
the $N$- and $\pi$-exchange FSI mechanisms;
in particular, 
the $\pi$-exchange, which was ignored in Ref.~\cite{tara-1},
plays a crucial role.
While no theoretical study 
has been done on FSI corrections for extracting $\gamma n\to\pi N$
polarization observables of the
current interests~\cite{clas1,mami2}, 
my analysis covers both unpolarized cross sections and polarization observables
$\Sigma$, $E$, and $G$~\cite{pol-def}.

The present analysis is based on 
a recently developed model for meson photoproduction off the
deuteron~\cite{gd-pinn-arxiv,etaN}.
The model includes, 
for both $d(\gamma,\pi^-)pp$ and $d(\gamma,\pi^0)pn$,
the impulse, $N$-exchange, and $\pi$-exchange mechanisms
as in previous 
investigations of $d(\gamma,\pi)NN$~\cite{arenhover,fix,sch10,wsl15}.
The ANL-Osaka model~\cite{knls13,knls16} is used
to generate the $\gamma N\rightarrow \pi N$
and $\pi N\rightarrow \pi N$ (off-shell) elementary amplitudes
that are built in the deuteron reaction model.
An update of 
$\gamma n\to N^*$ ($N^*$: bare nucleon resonance)
coupling parameters of the ANL-Osaka model
has been made by including 
recent data for $\gamma n\to\pi^-p$~\cite{clas1,clas2}
and $\gamma n\to\pi^0n$~\cite{mami1,mami2} in the fit~\cite{gd-pinn-arxiv}.
The initial deuteron wave function and 
half off-shell $NN\rightarrow NN$ amplitudes are 
those from the CD-Bonn potential~\cite{cdbonn}.
Comparisons of model predictions with data for $d(\gamma,\pi)NN$ are
presented in Ref.~\cite{gd-pinn-arxiv}.

For extracting cross sections of $\gamma n\to\pi N$ from those of
$d(\gamma,\pi)NN$, 
a formula that gives a relation between them is necessary.
For deriving it, 
one starts with the cross section formula for 
$\gamma(\bm{q})+ d(\bm{p}_d)\to \pi(\bm{k})+N_1(\bm{p}_1)+N_2(\bm{p}_2)$ 
[the laboratory-frame momenta are indicated in the parentheses]
as given by
\begin{eqnarray}
d\sigma_{\gamma d}&=&
(2\pi)^4
\delta^{(4)}({p}_d+{q}-{p}_1-{p}_2-{k})
\left[\frac{m_N}{E_{N_1}}
\frac{m_N}{E_{N_2}}\frac{1}{2E_\pi}\right]
\nonumber \\
&&\times 
\,|M_{f,i}(E)|^2\, 
\left[\frac{1}{2E_d} \frac{1}{2 E_\gamma}\right] 
d\bm{p}_1d\bm{p}_2d\bm{k} 
\ ,
\label{eq:dsigma-g}
\end{eqnarray}
where 
$E_x=\sqrt{\bm{p}_x^2+m_x^2}$ is the energy for
a particle $x$ with the momentum $\bm{p}_x$ and the mass $m_x$,
and detailed formulas for
the Lorentz invariant amplitude $M_{f,i}(E=E_\gamma+m_d)$ are given in 
Ref.~\cite{gd-pinn-arxiv}.
To isolate 
quasi-free events
by removing contributions from the other nucleon and FSI,
one conventionally applies a set of
kinematical cuts to the data. 
Within my calculation,
this amounts to restricting
the phase-space integral in Eq.~(\ref{eq:dsigma-g})
to obtain
${d^2\sigma_{\gamma d}(E_\gamma) / dW d\cos\theta} \rvert_{\rm cut}$
where, for $N_2$ being treated as a spectator,
$W$ is the invariant mass of the
final $\pi$-$N_1$ and $\theta$ the
angle between the momenta of $\gamma$ and $\pi$ in the
$\pi$-$N_1$ center-of-mass (CM) frame.
Then it is {\it assumed} that this partially integrated cross section is
solely from the quasi-free processes integrated over the same phase-space. 
The resulting formula that can be used in analyzing data is 
\begin{eqnarray}
{d^2\sigma_{\gamma d}(E_\gamma) \over dW d\cos\theta} \biggr\rvert_{\rm cut}
=
\phi(W;E_\gamma)\, {E'_\gamma\over E_\gamma}\,
{d\sigma_{\gamma n\to\pi N_1}(W) \over d\cos\theta} \ ,
\label{eq:dsig_extract_mod}
\end{eqnarray}
where $E'_\gamma$ is defined by
$E'_\gamma=(W^2-m^2_N)/(2 m_N)$ and thus is 
the photon energy in the laboratory frame for 
$\gamma n\to\pi N_1$.
The function $\phi(W;E_\gamma)$ gives the probability of finding
a process where the incident photon having
$E_\gamma$ hits a nucleon in the deuteron with an invariant
mass $W$, and is determined by the nucleon momentum distribution 
in the deuteron 
$\rho_d(p)\equiv u^2_s(p)+u^2_d(p)$
($u_{s,d}$: the deuteron $s,d$-wave radial function) as:
\begin{eqnarray}
\phi(W;E_\gamma) &=& \int\!\! d^3p_s {m_N\over E_N(\bm{p}_s)}\delta\left(
W-w(\bm{p}_s,E_\gamma) \right)
{\rho_d(|\bm{p}_s|)\over 4\pi}\nonumber\\
&&\times
\prod_i \theta(x_i^{\rm max}-x_i)\, \theta(x_i-x_i^{\rm min}),
\label{eq:eff_flux2}
\end{eqnarray}
with
\begin{eqnarray}
w (\bm{p}_s,E_\gamma) = \sqrt{ (E_\gamma+m_d-E_N(\bm{p}_s))^2
- (\bm{q}-\bm{p}_s)^2},
\label{eq:bar_W}
\end{eqnarray}
where $\bm{p}_s$
is the spectator nucleon momentum;
$\{x_i\}$ are a set of kinematical variables and 
$x_i^{\rm min}$ ($x_i^{\rm max}$) is the minimum (maximum) value allowed by
the cuts. 
Equations~(\ref{eq:dsig_extract_mod})-(\ref{eq:bar_W}) agree
with the formula presented in Ref.~\cite{tara,tara-1,laget-1,laget-2,laget-3}.
I also note that 
the relation
between the $\gamma d$ and $\gamma n$ cross sections in
Eqs.~(\ref{eq:dsig_extract_mod})-(\ref{eq:bar_W}) becomes exact
within my model
when considering only the `quasi-free' mechanism in which 
the incident photon interacts
with only one of the nucleons inside the deuteron, 
ignoring the FSI terms, crossed terms, and the small deuteron $d$-state.

Equation~(\ref{eq:dsig_extract_mod}) is the formula to extract
the $\gamma n\to \pi N_1$ cross section at a given $W$
from $d(\gamma,\pi)N_1N_2$ data where the $\pi N_1$ pair has the
invariant mass $W$.
This formula has been used in the MAMI
analyses~\cite{kossert,mami3,mami1,mami4,mami5,mami2,mami6}, but not in
recent JLab analyses~\cite{jlab1,clas1,clas2}.
The previous theoretical works~\cite{tara,tara-1}
presented a formula similar to Eq.~(\ref{eq:dsig_extract_mod}),
but did not use it either.
The formula practically used in the JLab and theoretical
analyses can be obtained 
by first assuming that $d\sigma_{\gamma n\to\pi N_1}(W) / d\cos\theta$ in 
the r.h.s. of Eq.~(\ref{eq:dsig_extract_mod}) is a constant in the range
of $W$ allowed by the kinematical cuts,
and then integrating both sides of Eq.~(\ref{eq:dsig_extract_mod}) over $W$.
The resulting formula is
\begin{eqnarray}
{d\sigma_{\gamma d}(E_\gamma) \over d\cos\theta} \biggr\rvert_{\rm cut}
&=& 
{d\sigma_{\gamma n\to\pi N_1}(\bar W) \over d\cos\theta}
\int d^3p_s\,
{m_N\over E_N(\bm{p}_s)}
{\rho_d(|\bm{p}_s|) \over 4\pi}
\nonumber\\
&&\times
{E'_\gamma\over E_\gamma}\,
\prod_i \theta(x_i^{\rm max}-x_i)\, \theta(x_i-x_i^{\rm min})
 \ ,
\label{eq:dsig_extract_mod2}
\end{eqnarray}
and $E'_\gamma/E_\gamma = E_N(\bm{p}_s)/m_N+\bm{p}_s\cdot\hat{q}/ m_N$.
The $\gamma n\to\pi N_1$ cross sections extracted with this formula
is an average over a certain range of $W$ with a certain weight
determined by $E_\gamma$, the Fermi motion, and the cuts.
The invariant mass $\bar{W}$ for this 
averaged cross section is usually identified with the one for the incident photon and
a nucleon at rest: $\bar{W}=\sqrt{2 m_N E_\gamma + m^2_N}$.
Equation~(\ref{eq:dsig_extract_mod2}) is thus based on the assumption that
the average (smearing) due to the Fermi motion does not
significantly invalidate this identification.
I will critically examine the validity of this formula.

\begin{table}[t]
\caption{
Kinematical cuts A, B, and C.
The cuts on the momenta for $\pi^-$ ($k_{\pi^-}$), 
faster proton ($p_f$), slower proton ($p_s$), 
and on the azimuthal angle difference between $\pi^-$ and the faster proton
$(\Delta\phi=|\phi_{\pi^-} - \phi_{p_f}|)$
in the 4-7th rows
have been used in the references listed in the second row
when extracting $\gamma n\to \pi^-p$ observables
 (specified in the parentheses) from $d(\gamma,\pi^-)pp$ data for 
the range of the photon energies listed in the third row.
}
\vspace*{-5mm}
\begin{center}
\renewcommand{\tabcolsep}{0.23cm}
\renewcommand{\arraystretch}{1.1}
\begin{tabular}[t]{cccc}\hline
& Cut A & Cut B & Cut C  \\
Ref. &\cite{jlab1} ($d\sigma/d\Omega$)& \cite{clas2} ($d\sigma/d\Omega$) & \cite{clas1} ($E$) \\
$E_\gamma$ (MeV) & 301 - 455 & 445 - 2510 & 700 - 2400\\\hline
$k_{\pi^-}$ (MeV)& $> 80$& $> 100$ & $> 400$  \\
$p_f$ (MeV)& $> 270$ & $> 360$ & $> 400$ \\
$p_s$ (MeV)& $< 270$ & $< 200$ & $< 100$  \\
$|\Delta\phi-180^\circ|$
& - & - & $< 20^\circ$  \\\hline
\end{tabular}
\end{center} 
\label{tab:1}
\end{table}
To proceed to a numerical study, 
the kinematical cuts need to be specified.
I choose realistic ones,
as summarized in Table~\ref{tab:1},
from recent JLab analyses~\cite{jlab1,clas1,clas2} 
where $\gamma n\to\pi^-p$ observables were extracted from
$d(\gamma,\pi^-)pp$ data.
In the previous theoretical studies~\cite{jlab1,tara,tara-1}, meanwhile,
simpler cuts were used. 
For extracting the $\gamma n\to\pi^0 n$ observable,
I apply the same cuts to $d(\gamma,\pi^0)pn$ 
by making obvious changes in the first column of
Table~\ref{tab:1}; `$\pi^-$' $\to$ `$\pi^0$', 
`Faster proton' $\to$ `Neutron', and 
`Slower proton' $\to$ `Proton'.
When extracting the $\gamma n\to\pi N$ observables in this paper,
I use Cut A (Cut B) for $E_\gamma = 300$~(500)~MeV.
For $E_\gamma>$ 700~MeV, 
Cut B (Cut C) is used for extracting unpolarized
differential cross sections 
(polarization asymmetries $\Sigma$, $E$, and $G$). 
I always take these choices for the cuts, 
unless otherwise stated.

To implement any kinematical cuts into the numerical computations, 
it is most convenient to use the Monte-Carlo method when performing
the phase-space integrals in Eqs.~(\ref{eq:dsigma-g}),
(\ref{eq:eff_flux2}),
and (\ref{eq:dsig_extract_mod2}).
The bin sizes (resolutions) for $\cos\theta$ and $W$
are taken to be 0.1 and
10~MeV, respectively, and thus a numerical value at
$x$ (= $\cos\theta$, $W$) is understood to be
the average over the range of $x-\Delta/2 \le x \le x+\Delta/2$ where
$\Delta$ is the bin size.
Therefore, all numerical results 
(except those for free $\gamma n\to \pi N$ observables)
are given with central values and their statistical errors associated with the Monte-Carlo method. 

\begin{figure}
  \includegraphics[width=.5\textwidth]{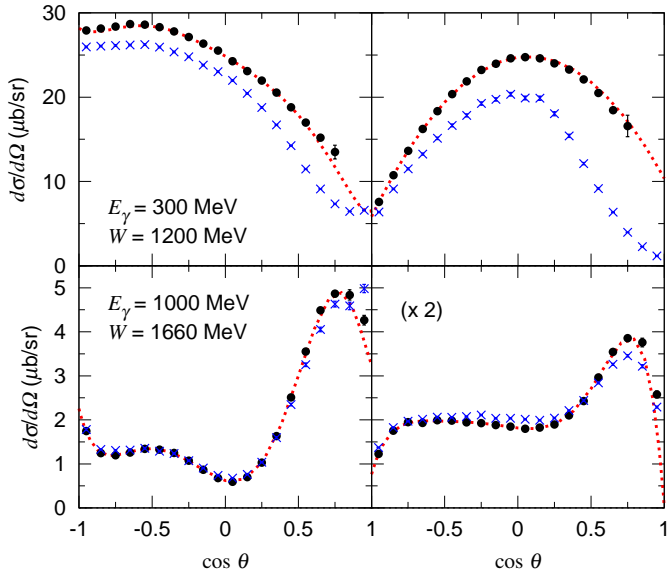}
\caption{
Unpolarized differential cross sections for $\gamma n\to\pi^-p$ (left) and 
$\gamma n\to \pi^0n$ (right) in the CM frame extracted from 
$d(\gamma,\pi^-)pp$  and
$d(\gamma,\pi^0)pn$ data
generated from my model including only the quasi-free mechanism;
$E_\gamma=300\ (1000)$~MeV for 
the upper (lower) row.
The black circles [blue crosses]
are extracted using 
Eq.~(\ref{eq:dsig_extract_mod}) [Eq.~(\ref{eq:dsig_extract_mod2})], 
and $W$ in
Eq.~(\ref{eq:dsig_extract_mod}) is indicated in the panels.
The errors are only statistical from the Monte-Carlo integral, and are
not shown when smaller than the point size.
The red dotted curves are the free $\gamma n\to \pi N$ cross sections
at $W$ from the ANL-Osaka model.
The cross sections are scaled by the factor in the parenthesis when it
is given.
}
\label{fig:extracted-QF}
\end{figure}

\begin{figure}
  \includegraphics[width=.5\textwidth]{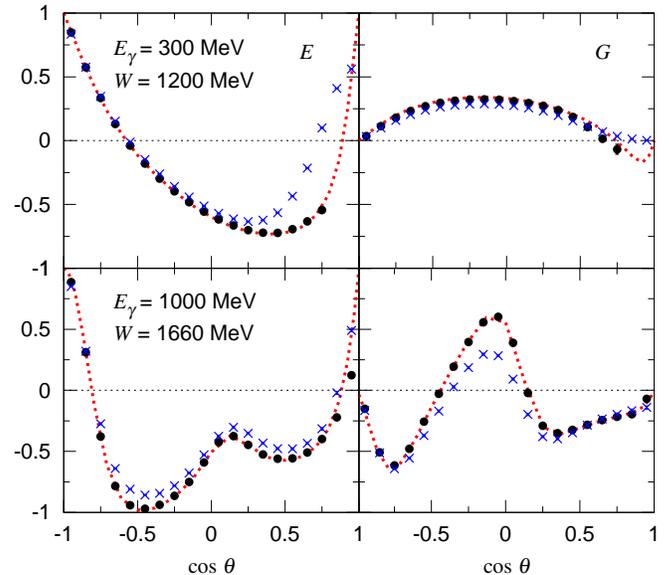}
\caption{
The polarization observables 
$E$ (left) and $G$ (right)
for $\gamma n\to \pi^-p$
extracted from $d(\gamma,\pi^-)pp$.
For $E_\gamma = 300$~(1000)~MeV,
Cut A (Cut B) of Table~\ref{tab:1} is used.
The other features are the same as those in Fig.~\ref{fig:extracted-QF}.
}
\label{fig:extracted-QF2-3}
\end{figure}

My investigation goes as follows. For a given choice of the kinematical cuts, 
my model generates
${d^2\sigma_{\gamma d}(E_\gamma) / dW d\cos\theta} \rvert_{\rm cut}$
[${d\sigma_{\gamma d}(E_\gamma) / d\cos\theta} \rvert_{\rm cut}$]
as `data' from which 
$d\sigma_{\gamma n\to\pi N}(W\, [\bar W]) / d\cos\theta$
is extracted with Eq.~(\ref{eq:dsig_extract_mod})
[Eq.~(\ref{eq:dsig_extract_mod2})].
By comparing the extracted $d\sigma_{\gamma n\to\pi N}(W\, [\bar W]) / d\cos\theta$
with the corresponding one on a free neutron, which is calculated with
the same elementary amplitudes used in calculating
the $\gamma$-$d$ cross sections,
I can examine the extent to which 
the extracted cross sections are distorted by
the  FSI and/or the Fermi motion.

\begin{figure*}[t]
 \includegraphics[width=1\textwidth]{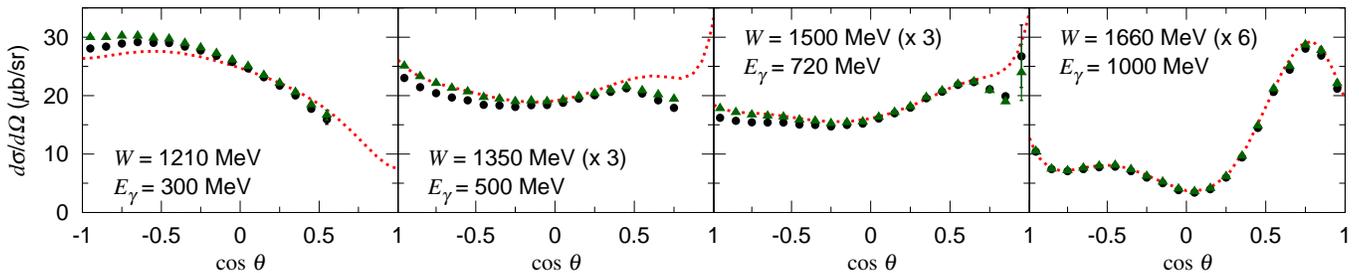}
\caption{
The pion angular distribution for $\gamma n\to \pi^-p$.
The black circles (green triangles) are extracted 
using Eq.~(\ref{eq:dsig_extract_mod}) from
$d(\gamma,\pi^-)pp$ generated by
my model including the impulse + $N$-exchange + $\pi$-exchange 
(impulse + $N$-exchange) terms.
The other features are the same as those in
Fig.~\ref{fig:extracted-QF}.
}
\label{fig:gn-pimp-DC}
\end{figure*}
I first confirm in Fig.~\ref{fig:extracted-QF} 
[Fig.~\ref{fig:extracted-QF2-3}] that
the extracted $\gamma n\to\pi^0n,\pi^-p$ unpolarized cross sections 
[$E, G$] shown by the black circles
reproduce the corresponding free
ones given by the red dotted curves,
when Eq.~(\ref{eq:dsig_extract_mod}) is used for the extraction
and the l.h.s. is calculated 
with the quasi-free mechanism only. 
The $\gamma$-$n$ observables at 
$W=\bar{W}= \sqrt{2 m_N E_\gamma + m^2_N}\simeq$~1200 (1660)~MeV 
are extracted 
from $\gamma$-$d$ at $E_\gamma =$ 300 (1000)~MeV.
Cut A (Cut B) of Table~\ref{tab:1} is used for 
$E_\gamma = 300$~(1000)~MeV.
The absence of the black circles 
in the forward pion angles for $E_\gamma=300$~MeV
is because these
pion angles are not allowed by the kinematical cuts including the cut on $W$.

In the same figures, the blue crosses represent
the $\gamma n\to\pi^0n,\pi^-p$ observables 
extracted with Eq.~(\ref{eq:dsig_extract_mod2}).
Because Eq.~(\ref{eq:dsig_extract_mod2}) does not include the $W$-cut, 
the extracted $\gamma$-$n$ observables are an average over a
range of $W\sim 1180-1210$ ($1600-1700$)~MeV for $E_\gamma=300$ (1000)~MeV.
These $\gamma$-$n$ observables
are clearly different from
the corresponding free ones in some cases.
In particular, the extracted cross sections for $\bar W=1200$~MeV
are significantly smaller than the free ones.
This difference can be explained as follows.
The $\gamma n \rightarrow \pi N$  
cross sections in the $W= 1180-1210$ MeV region change
rapidly and reach the $\Delta(1232)$ peak near $W\simeq 1200$~MeV.
Because the blue crosses in Fig.~\ref{fig:extracted-QF}
are the average 
of $d\sigma_{\gamma n\to\pi N}(W)/d\cos\theta$
over this range of $W$ 
(average of a $W<1200$~MeV region for $\cos\theta \gtap 0.7$),
they are
necessarily smaller than the free ones at $W=1200$~MeV.

At $E_\gamma=1000$~MeV ($\bar{W}=1660$~MeV),
on the other hand, 
the blue crosses in Fig.~\ref{fig:extracted-QF} reproduce 
the free cross sections fairly well
in the $\cos\theta <  0.5$ region.
This is because the free cross sections in the range of
$W\simeq 1600-1700$~MeV has a 
mild and monotonic
$W$-dependence, and hence the average is not significantly
different from the free one at $W=\bar{W}$. 
In $\cos\theta > 0.5$, however,
the average does not
cancel out the $W$-dependence very well,
giving the extracted cross sections visibly different from
the free ones.

Differences between the blue crosses and red dotted curves 
are also seen in Fig.~\ref{fig:extracted-QF2-3}
for the polarization asymmetries of $\gamma n\to\pi^-p$,
such as $G$ at $\bar W=1660$~MeV.
The difference again stems from averaging 
a rapid and non-monotonic $W$-dependence of $G$ for the free 
$\gamma n\to\pi^-p$ around $W=1660$~MeV.
The asymmetry $E$ at $\bar W=1200$~MeV
also significantly deviates from the free one in $\cos\theta \gtap 0.3$.
In these pion angles, $\gamma n\to\pi^- p$ at $W\sim 1200$~MeV in the deuteron
are largely eliminated by the kinematical cuts, as implied by the
absence of the black circles in $\cos\theta \gtap 0.6$,
and $\gamma n\to\pi^- p$ of $W\sim 1120-1180$~MeV mainly contribute here.
The average over this range of $W$ gives the blue crosses
which significantly deviate from the free $E$ of $W=1200$~MeV.
Thus Figs.~\ref{fig:extracted-QF} and \ref{fig:extracted-QF2-3}
indicate that the neutron-target observables extracted with
Eq.~(\ref{eq:dsig_extract_mod2}) can seriously suffer from the Fermi
smearing, even when using the kinematical cuts of Table~\ref{tab:1} where
the cut on the spectator momentum ($p_s$) should limit the $W$-range.
One can avoid this problem by using
Eq.~(\ref{eq:dsig_extract_mod}) 
that includes the $W$-cut.

Now I study FSI effects on neutron-target observables extracted
using Eq.~(\ref{eq:dsig_extract_mod}).
I calculate $d(\gamma,\pi^-)pp$ cross sections including 
the impulse, $N$-exchange, and $\pi$-exchange (impulse and
$N$-exchange) terms, apply the kinematical cuts, and extract
$\gamma n \rightarrow \pi^-p$ unpolarized differential cross sections
as shown by the black circles (green triangles)
in Fig.~\ref{fig:gn-pimp-DC}.
The differences between these results and the 
free cross sections (red dotted curves) indicate 
that the extracted $d\sigma_{\gamma n\to\pi^-p}(W)/d\Omega$ 
contain some FSI effects
even after the kinematical cuts have been applied.
At $E_\gamma = 500$ and 720~MeV,
the $N$-exchange ($\pi$-exchange) FSI effect is to visibly reduce the cross
sections in the forward (backward) pion region, which is qualitatively consistent
with the findings in Ref.~\cite{tara}.
Meanwhile, at $E_\gamma = 1000$~MeV, 
a reduction due to the $N$-exchange in the forward pion angles
is canceled by an enhancement due to
the impulse crossed term.

\begin{figure}
  \includegraphics[width=.5\textwidth]{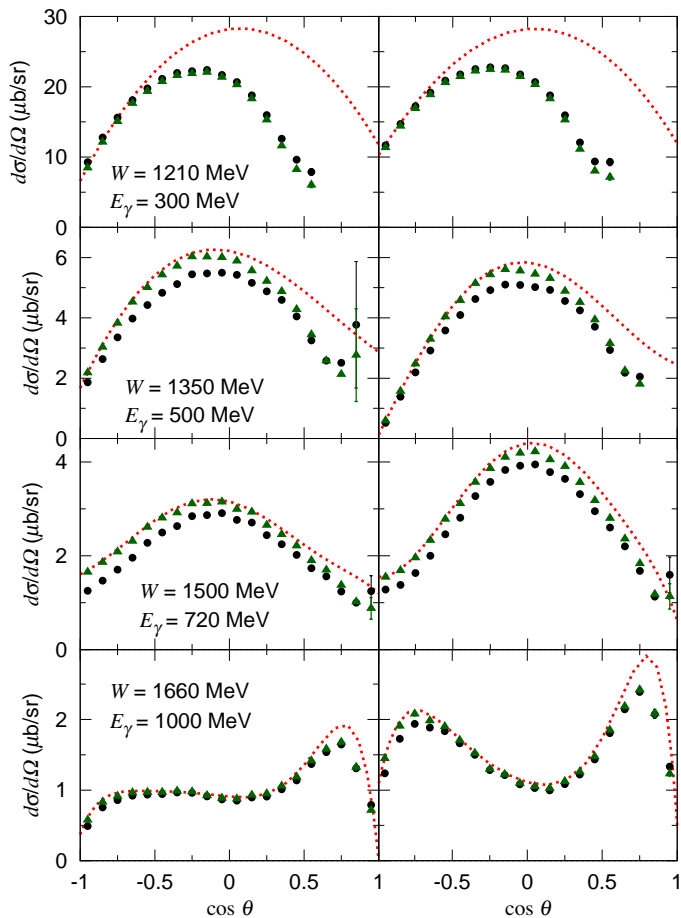}
\caption{(Left [Right])
The pion angular distribution for 
$\gamma n\to \pi^0n$ 
[$\gamma p\to \pi^0p$] 
extracted from 
$d(\gamma,\pi^0)pn$.
The other features are the same as those in Fig.~\ref{fig:gn-pimp-DC}.
}
\label{fig:gn-pi0n-DC}
\end{figure}

The unpolarized $\gamma n\to\pi^0 n$ differential cross
sections extracted from $d(\gamma,\pi^0)pn$ using Eq.~(\ref{eq:dsig_extract_mod})
are shown in Fig.~\ref{fig:gn-pi0n-DC} (left).
FSI effects are clearly larger
than in the case of $\gamma n\to\pi^- p$.
The $N$-exchange FSI largely reduce the cross sections 
at $E_\gamma = 300$ MeV of the $\Delta(1232)$ region.
This is because 
the deuteron component (coherent process) is included in
the $NN$ plane wave from the impulse mechanism, and is
eliminated by the scattering $NN$ $^3S_1$ partial wave thanks to the
orthogonality~\cite{arenhover,fix}. 
Although the reduction due to the $N$-exchange FSI becomes smaller 
as the photon energy increases, it still persists in the forward pion region.
Meanwhile, the $\pi$-exchange FSI effect is negligibly small at
$E_\gamma = 300$ MeV, as indicated by 
the small differences between 
the black circles and green triangles.
As the photon energy increases, however, 
the $\pi$-exchange FSI
significantly reduces the cross sections overall except the forward
pion angles.
On the other hand, in Ref.~\cite{tara-1} where FSI corrections for
extracting $\gamma n\to\pi^0 n$ cross sections were theoretically
estimated, 
the authors did not consider the $\pi$-exchange mechanism, assuming its
effect negligible. 
As a result, the estimated FSI effects are only from 
the $N$-exchange mechanism and are
visible only in the forward pion kinematics 
($\cos\theta\gtap 0.85$ for $E_\gamma= 787$~MeV, for example).

\begin{figure}
  \includegraphics[width=.5\textwidth]{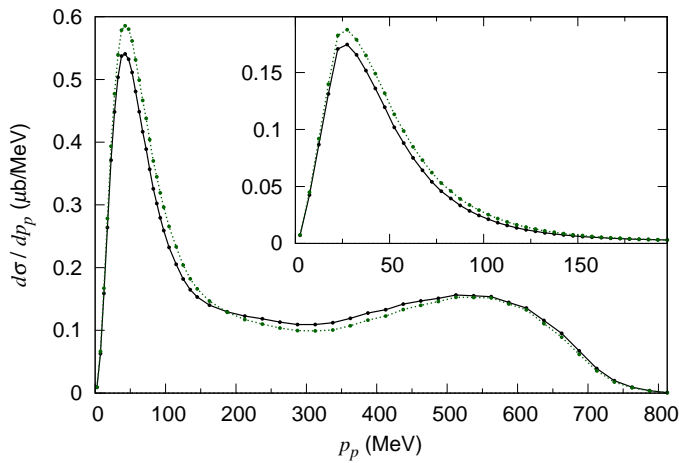}
\caption{
Proton momentum ($p_p$) distribution for
$d(\gamma,\pi^0)pn$ at $E_\gamma=500$~MeV.
The points connected by the black solid (green dotted) lines are from 
my model including the impulse + $N$-exchange + $\pi$-exchange 
(impulse + $N$-exchange) terms.
Insert: Same distribution but 
Cut~B ($p_s=p_p$, $p_f=p_n$)
and the $W$-cut ($|W-1350|<5$~MeV) have been applied.
}
\label{fig:gn-pi0n-ps}
\end{figure}
The pronounced $\pi$-exchange FSI effect would call for an
explanation because previous
calculations~\cite{gd-pinn-arxiv,arenhover,fix,wsl15}
showed it to be rather small for $d(\gamma,\pi^0)pn$.
In the proton momentum distribution of $d(\gamma,\pi^0)pn$
as shown in Fig.~\ref{fig:gn-pi0n-ps},
the difference between the black solid and green dotted lines
represents the $\pi$-exchange FSI effect.
Without kinematical cuts,
the $\pi$-exchange FSI reduces (enhances) the spectrum for 
$p_p \ltap 190$ ($p_p \gtap 190$)~MeV.
The net $\pi$-exchange effect is small 
after integrating over $p_p$, 
and the previous calculations found this small effect.
When the kinematical cuts are applied,
the (spectator) proton momentum distribution becomes to the one shown in
the insert.
Because of cutting the large spectator momentum, 
the enhancement due to the $\pi$-exchange has been removed and only the
reduction in the small spectator momentum region remains.
This reduction appears as
the significant $\pi$-exchange FSI effect 
in Fig.~\ref{fig:gn-pi0n-DC}.

The $\gamma p\to \pi^0 p$ cross sections 
have been also extracted from
$d(\gamma,\pi^0)pn$ and shown in Fig.~\ref{fig:gn-pi0n-DC} (right).
Comparing with Fig.~\ref{fig:gn-pi0n-DC} (left),
one can see that 
the FSI-induced reduction factors $R_{\rm FSI}$, defined by
the black circles divided by the red dotted curves
(thus also including effects from the impulse crossed term),
for $\gamma n\to\pi^0 n$ and $\gamma p\to\pi^0 p$
are within a few percents difference in most cases.
This may partially support the assumption
 in the MAMI analysis~\cite{mami1,mami6} 
that the FSI effects are the same for both.
However, the FSI effects are sometimes more different.
In the third row of Fig.~\ref{fig:gn-pi0n-DC} and $\cos\theta\sim -1$, for example,
$R_{\rm FSI}\sim 0.74$ (0.81)
for $\gamma n\to\pi^0 n$ ($\gamma p\to\pi^0 p$).
The MAMI analysis~\cite{mami1,mami6} obtained 
the experimental counterpart to $R_{\rm FSI}$, denoted by $R^{\rm exp}_{\rm FSI}$,
using their 
$\gamma p\to\pi^0 p$ cross section data measured on hydrogen and deuterium
targets.
One can find that $R_{\rm FSI}$ for $\gamma p\to\pi^0 p$
from my calculation is 
qualitatively very similar to $R^{\rm exp}_{\rm FSI}$.
For a more quantitative comparison, 
the same kinematical cuts 
as in Refs.~\cite{mami1,mami6}
should be used because the FSI effects can depend on the choice of cuts. 
Still, my model explains a major fraction of $R^{\rm exp}_{\rm FSI}$ for the
first time.

It has been clearly shown that unpolarized differential cross
sections for $\gamma n \rightarrow \pi N$ extracted with
Eq.~(\ref{eq:dsig_extract_mod})
and the kinematical cuts listed in Table~\ref{tab:1}
need to be appropriately corrected for the FSI.
It is however not trivial to develop a
formula giving the necessary correction factors $R^{-1}_{\rm FSI}$,
since the corrections
strongly depend on $E_\gamma$, $W$, $\theta$ (pion angle), and cuts.
The use of a dynamical model, such as used in this paper, to calculate
the corrections for each data analysis is perhaps necessary in practice.
A final remark is that
the extracted polarization asymmetries $\Sigma$, $E$, and $G$ 
for $\gamma n \rightarrow \pi N$ are
numerically confirmed to be 
reasonably safe from distortions caused by the FSI,
provided that the extraction is done with Eq.~(\ref{eq:dsig_extract_mod}).

\begin{acknowledgments}
The author thanks H. Kamano, T.S.-H. Lee, and T. Sato for useful discussions.
The author also thanks A. Sandorfi, T. Kageya, D. Carman, B. Krusche, and M. Dieterle for
useful information on their experimental data and encouragements.
This work is in part supported by 
Funda\c{c}\~ao de Amparo \`a Pesquisa do Estado de S\~ao Paulo (FAPESP),
Process No.~2016/15618-8.
Numerical computations in this work were carried out
with SR16000 at YITP in Kyoto University,
the High Performance Computing system at RCNP in Osaka University,
the National Energy Research Scientific Computing Center, which is
supported by the Office of Science of the U.S. Department of Energy
under Contract No. DE-AC02-05CH11231, 
and the use of the Bebop [or Blues] cluster in the Laboratory Computing
Resource Center at Argonne National Laboratory.
\end{acknowledgments}



\end{document}